\documentclass[aps, prd, groupedaddress, eqsecnum, nofootinbib, showpacs, preprintnumbers]{revtex4}

\usepackage{latexsym}
\usepackage{amsfonts}
\usepackage{amsmath}
\usepackage{amsthm}
\usepackage{amssymb}
\usepackage{amsbsy}
\usepackage{bbm}
\usepackage[latin9]{inputenc}
\usepackage{nicefrac}
\usepackage[stable]{footmisc}
\usepackage{afterpage}
\usepackage{ifpdf}
\ifpdf
\usepackage[pdftex,unicode,implicit]{hyperref}
\hypersetup{%
  pdftitle    = {Born-Infeld with Higher Derivatives},
  pdfkeywords = {electromagnetism, duality, non-linear, selfduality},
  pdfauthor   = {Wissam Chemissany, Renata Kallosh, Tomas Ortin},
  plainpages  = true,
  colorlinks  = true,
  citecolor   = blue,
  urlcolor    = red,
  linkcolor   = black
}
\newcommand{\hepth}[1]{arXiv:{\tt
\href{http://www.arXiv.org/abs/hep-th/#1}{hep-th/#1}}}

\newcommand{\arxiv}[1]{{\tt
\href{http://www.arXiv.org/abs/#1}{arXiv:#1}}}
\else
  \usepackage[dvips]{graphicx}
  \usepackage[unicode,implicit]{hyperref}
  \newcommand{\hepth}[1]{arXiv:{\tt hep-th/#1}}

  \newcommand{\arxiv}[1]{{\tt arXiv:#1}}
\fi

\begin{document}

\preprint{SU-ITP-11/50}
\preprint{IFT-UAM/CSIC-11-92}

\title{Born-Infeld with Higher Derivatives}

\author{\bf Wissam Chemissany${}^a$, Renata Kallosh${}^b$, 
and Tomas Ortin${}^c$}

\affiliation{\vskip .43cm 
  ${}^a$ Department of Physics and Astronomy,
  University of Waterloo,
  Waterloo, Ontario, Canada, N2L 3G1\\
  ${}^b$Stanford Institute for Theoretical Physics and Department of Physics,
  Stanford University,
  Stanford, CA 94305-4060, USA\\
  ${}^c$ Instituto de Fisica Teorica UAM/CSIC, C/ Nicolas Cabrera, 13-15,
  C.U. Cantoblanco, E-28049-Madrid, Spain }

\begin{abstract}
  We present new models of non-linear electromagnetism which satisfy the
  Noether-Gaillard-Zumino current conservation and are, therefore,
  self-dual. The new models differ from the Born-Infeld-type models in that
  they deform the Maxwell theory starting with terms like $\lambda (\partial
  F)^{4}$.  We provide a recursive algorithm to find all higher order terms in
  the action of the form $\lambda^{n} \partial ^{4n} F^{2n+2} $, which are
  necessary for the $U(1)$ duality current conservation. We use one of these
  models to find a self-dual completion of the $\lambda (\partial F)^{4}$
  correction to the open string action.  We discuss the implication of these
  findings for the issue of UV finiteness of ${\cal N}=8$ supergravity.
\end{abstract}


\maketitle

\pagestyle{plain}
\setcounter{page}{1}


\section{Introduction}

In this paper we discuss a method for constructing effective Lagrangians for
non-linear theories with duality symmetries. This work builds on earlier
papers by \cite{Kallosh:2011dp}, \cite{arXiv:1105.1273},
\cite{Carrasco:2011jv}. The hope is that this procedure may shed further light
on counterterms in maximal supergravity theories. In particular it may improve
our understanding of the role of $E_{7(7)}$ electro-magnetic duality symmetry
in ${\cal N}=8$ supergravity.

Here we study a simplified class of models with only one vector field, no
scalars and duality group $U(1)$.  Although the $E_{7(7)}$ symmetry of ${\cal
  N}=8$ supergravity is a global continuous symmetry it has some unusual
features which were uncovered for the first time in 1981 by Gaillard and
Zumino \cite{Gaillard:1981rj} in the construction of extended supergravities
(for a recent review see \cite{Aschieri:2008ns}). The familiar global
continuous symmetries are defined by the Noether current conservation and are
well known since 1918. However, duality symmetries have subtleties in the
vector sector of the theory. Namely, the vector part of the action is not
invariant under duality symmetry, but transforms in a specific way, so that
the Bianchi identities and equations of motion transform into each other by
duality symmetry. This feature is guaranteed by the conservation of the
Noether-Gaillard-Zumino current and the corresponding NGZ identity.

Several theories with $U(1)$ duality are known. At the free, linear, level,
there is Maxwell's electromagnetism and the higher-derivative generalizations
constructed in \cite{arXiv:1105.1273}. At the interacting, non-linear, level,
there is the Born-Infeld (BI) theory \cite{BI,Schrodinger,LEBEDEV-85-193} and
its generalizations \cite{Gibbons:1995cv,Gaillard:1997rt}. The fact that the
original BI theory has electromagnetic duality was first noticed by
Schr\"{o}dinger \cite{Schrodinger}. The action of this model and of the
generalizations constructed so far only contain powers of the Maxwell field
strength $F$, and no higher derivatives.  The BI Lagrangian had been derived
by Fradkin and Tseytlin \cite{LEBEDEV-85-193} as the low-energy spacetime
effective Lagrangian for the vector field with a constant field strength,
coupled to a string. The self-duality of Born-Infeld action and the relation
to 
the D3-brane of type IIB superstring theory and its SL(2, Z)-symmetry was
studied in \cite{hep-th/9602064}.  For a review on BI action and open
superstring theory we refer to \cite{hep-th/9908105}.

The action of the BI model has a well-known closed form
$\det^{1/2}(\eta_{\mu\nu} + F_{\mu\nu})$, while the actions of its
generalization do not, so the Lagrangian has to be written as an infinite
power series. Gibbons and Rasheed \cite{Gibbons:1995cv} have shown that there
is a function of one variable's worth of Lagrangians admitting duality
rotations and gave an explicit algorithm for their construction. These models
were developed in more detail in \cite{Gaillard:1997rt} and more recently in
\cite{Carrasco:2011jv}. The action of all these models is identical at the
$F^{2}, F^{4}, F^{6}$ level, but they differ at the $F^{8}$ and higher levels.

In this paper we will construct two simple self-dual models of non-linear
electrodynamics whose first deviation from the free Maxwell theory starts with
a $(\partial F)^{4}$ term and contain terms of higher order in $F$ and
derivatives. We will present recursive procedure to construct all of them.

A term of this kind $\Big ( (\partial F)^{4}\Big )$ is known to arise in the
4-point amplitude of the open string\footnote{The same type of terms have been
  considered in \cite{Green:2000ke} as part of the effective action of a
  single D3-brane. They have been shown to fit elegantly into an
  $\textrm{SL}(2,\mathbbm{Z})$-invariant function that encodes both the
  perturbative and non-perturbative contributions to the amplitude.}
\cite{Print-88-0247 (LEBEDEV)}.  It was shown in \cite{Chemissany:2006qd}
that, with this term (and other $F^{4}$ with higher derivatives present in the
4-point amplitudes), the theory satisfies the NGZ identity, and is consistent
with electro-magnetic self-duality. Here we will show that a combination of
the two simple self-dual models gives precisely the $(\partial F)^{4}$ term
studied in \cite{Chemissany:2006qd} as well as higher-order terms required to
satisfy the NGZ current conservation at the $n$-point level.

We will also describe a more general class of models where there are terms
with $F^{n}$, without derivatives, as well as terms with derivatives
$\partial^{2m} F^{2n}$. In all cases the algorithm for a construction of such
actions satisfying the NGZ identity will be given.


\section{$U(1)$  Duality, No Scalars}
\label{dualityEM}

Our goal is to construct actions $S(F)$, where
$F_{\mu\nu}\equiv \partial_{\mu} A_{\nu}-\partial_{\nu} A_{\mu}$ is the
Maxwell field strength, which have a non-linear $U(1)$ duality.  Two classes
of such actions are known in the literature: that of the Born-Infeld theory
and its generalizations \cite{Gibbons:1995cv,Gaillard:1997rt}, that depend
only on $F$ and not on its derivatives, and the action constructed in
\cite{arXiv:1105.1273} which has higher derivatives but is quadratic in $F$.

As usual, we define the dual field strength $G(F)$ by 
\begin{equation}
\label{covConstraint}
\tilde{G}^{\mu \nu} \equiv \tfrac{1}{2}\epsilon^{\mu\nu\rho\sigma} 
G_{\rho\sigma}
\equiv 2 \frac{ \delta  S(F)}{\delta  F_{\mu \nu}}\, .
\end{equation}
The infinitesimal $U(1)$ duality transformations that interchange the
equations of motion ${\partial}_{\mu} \tilde{G}^{\mu\nu}=0$ and Bianchi
identities ${\partial}_{\mu} {\tilde F}^{\mu\nu} =0$ are given by
\begin{equation} 
\label{constraintMaxwell}
\delta 
\left(
  \begin{array}{cc}
    F \\
    G \\
  \end{array}
\right)\ =\left(
  \begin{array}{cc}
    0 &  B \\
    -B  &0 \\
  \end{array}
\right)  \left(
  \begin{array}{cc}
    F \\
    G \\
  \end{array}
\right) \, .  
\end{equation} 
The necessary condition for the theory to be selfdual is conservation of the
the NGZ current \cite{Gaillard:1981rj}, which in $U(1)$ models without scalars
requires that
\begin{equation} 
 \label{consistencyEqn}
\int d^{4}x ( F\tilde{F} + G\tilde{G} )=0\, .
 \end{equation}
 The $U(1)$ case is a special case of a more general $Sp(2n, \mathbb{R})$
 duality group
 \begin{equation} 
\label{constraint}
\delta 
\left(
  \begin{array}{cc}
    F \\
    G \\
  \end{array}
\right)\ =\left(
  \begin{array}{cc}
    A &  B \\
    C  & D \\
  \end{array}
\right)  \left(
  \begin{array}{cc}
    F \\
    G \\
  \end{array}
\right) \, . 
\end{equation}
which also acts on scalars, $\delta \phi =\delta \phi (A,B,C,D)$. 

In the general case, the NGZ identity requires the action to be of the form
\cite{Gaillard:1981rj}
\begin{equation}  
\label{actionForm}
S=\tfrac{1}{4} \int d^{4}x \, F
{\tilde G} +S_{\rm inv}\, ,
\end{equation}
where $S_{\rm inv}$ is exactly invariant under the duality group $\delta
S_{\rm inv}=0$.  This is a reconstructive identity, since, in principle, it
may be used to find the action from the knowledge of $S_{\rm inv}$ and $G(F)$.
On the other hand,
\begin{equation} 
\label{linearvar}
\delta S= \tfrac{1}{2} \int  d^{4}x  \, \delta ( F\tilde G )= \tfrac{1}{4}
\int d^{4}x (\tilde G B G+ \tilde FC F) \, .
\end{equation}
Eqs.~(\ref{actionForm}) and (\ref{linearvar}) are equivalent to NGZ current
conservation, whereas eq.  (\ref{consistencyEqn}) is a particular form of the
current conservation, valid only for $U(1)$ models without scalars. Indeed,
only for $U(1)$ in absence of scalars $\delta S= \tfrac{1}{2} \tilde GBG$, with
$B=-C$ and $A=0$ and therefore
\begin{equation}  
\tfrac{1}{2}
\tilde GBG= \tfrac{1}{4} ( \tilde G B G+ \tilde FC F) \qquad \Rightarrow \qquad
\int d^{4}x ( F\tilde{F} + G\tilde{G} )=0 \ .  
\end{equation}


\subsection{New reconstructive identity in $U(1)$ models 
without scalars}

As mentioned before, to use the generic reconstructive identity
(\ref{actionForm}) one needs, in addition to $G(F)$ in each particular model,
additional information on $S_{\rm inv}$. In the $U(1)$ models without scalars
that we are considering here, this additional information comes form the
following general observation: if $\lambda$ is the coupling constant of the
model (so the linear Maxwell term is independent of it), then, $S_{\rm inv}$
is related to the full action by
\begin{equation}
\label{observation}
S_{\rm inv} = -\lambda \frac{\partial S}{\partial \lambda}\, .  
\end{equation}
This relation follows from the uniqueness of $S_{\rm inv}$ for a given
non-linear theory, with given non-linear duality transformations and from the
invariance of $\lambda \frac{\partial S}{\partial \lambda}$. The precise
coefficient relating these two objects follows from the study of the linear
and next-to-linear terms of a generic action. For example, it is well known
\cite{Gaillard:1981rj} that in the BI model one has $S_{\rm inv}= -g^{2}
{\partial S\over \partial g^{2}}$.

Using this general observation, one can derive a new, more useful,
reconstructive identity:

\begin{equation}  
\label{new}
S(F)= {1\over 4 \lambda} \int d^{4}x \, d\lambda \, F\tilde G \, .
\end{equation}

To prove is, we will first prove that $\lambda {\partial S\over \partial
  \lambda}$ is duality invariant\footnote{This is just a particular case of
  the general theorem proven in Appendix~B of
  Ref.~\cite{Gaillard:1981rj}. Note that the general proof in
  Ref.~\cite{Gaillard:1981rj} is based on a condition that the duality
  transformation of scalars do not depend on a coupling associated with the
  deformation, meanwhile, the transformation law of vectors does depend on
  such a coupling. This raises the issue whether in extended supersymmetric
  theories, where scalars and vectors are in the same multiplet, the
  construction of this type is available.}, using the NGZ identity and the
definition (\ref{covConstraint}) (\ref{consistencyEqn}):
\begin{equation} 
  \lambda \frac{\partial}{\partial \lambda} \int d^{4}x (F\tilde{F}+G\tilde{G})
  =
  2  \lambda \int d^{4}x \frac{\partial \tilde G}{\partial \lambda} G
=
 \lambda \int d^{4}x \frac{\partial}{\partial \lambda}
\left(\frac{\delta S}{\delta F}\right) G  
=0 \, .
\end{equation}
Then, since the functional variation and the partial derivative with respect
to $\lambda$ commute, and using  (\ref{constraintMaxwell}), we find that 
\begin{equation} 
0
=
\int d^{4}x 
\frac{\delta }{\delta F}
\left(\lambda \frac{\partial S}{\partial \lambda}
\right) G  
=
B^{-1}
\int d^{4}x 
\frac{\delta }{\delta F}
\left(\lambda \frac{\partial S}{\partial \lambda}
\right) \delta F
= 
B^{-1}
\delta \left(\lambda
  \frac{\partial S}{\partial\lambda}\right)\, .
\end{equation}
Now, using the observation (\ref{observation}) in (\ref{actionForm}) that
\begin{equation} 
S 
+\lambda {\partial S\over \partial \lambda}
= 
\tfrac{1}{4} \int d^{4} x\,   F\tilde G \, ,
\end{equation}
which can integrated immediately, leading to (\ref{new}).

The new reconstructive identity (\ref{new}) is particularly well-suited to
find the action as a series expansion in $\lambda$ when the dual field
strength $G$ is also available as a series expansion in $\lambda$:
defining\footnote{The global factors of $2$ have been introduced for later
  convenience, since they lead to simpler expressions for the coefficients
  $T^{(n)\pm}$ to be introduced later.}
\begin{eqnarray}
\label{eq:Gexpansion}
\tilde{G}(F) 
& = &   
-F+
2 \sum_{n=1}^{\infty}\lambda^{n}\tilde{G}^{(n)}(F)\, ,
\\
& & \nonumber \\
\label{eq:Sexpansion}
S
& = &   
-\tfrac{1}{2}\int \, d^{4}x \, F^{2}
+
2 \sum_{n=1}^{\infty}\lambda^{n}S^{(n)}\, ,
\end{eqnarray}
so that  the $\lambda=0$ free limit of the theory is the Maxwell theory,
we find that each term in the expansion of the action is given by 
\begin{equation}
S^{(n)}=\frac{1}{4(n+1)}\int d^{4}x\, F\tilde{G}^{(n)}(F)\, .   
\end{equation}

In the models that we are going to consider $\tilde{G}(F)$ is given by a
series expansion of the above form with all the terms of higher order in
$\lambda$ given by a simple recursion relation and these results can be
checked explicitly order by order in $\lambda$.

  
\subsection{NGZ identity with graviphoton convention}

To proceed, we introduce the standard supergravity graviphoton conventions
\cite{Andrianopoli:1996ve}, employed in \cite{Carrasco:2011jv} in the
covariant procedures for perturbative non-linear deformation of
duality-invariant theories. In the complex basis we define
\begin{equation}
  \label{complexb} 
T=F-iG,\qquad T^{*}=F+iG  \, ,
\end{equation}
which transform under finite $U(1)$ duality transformations with a phase, so,
under (\ref{constraintMaxwell})
\begin{equation}
\delta T = iBT\, .  
\end{equation}
We also introduce the self-dual notation,
\begin{equation} 
\label{selfDualT}
T^\pm= \textstyle{ \frac{1}{ 2}}(T\pm i \tilde T)\, . 
\end{equation}
and form 4 different combinations of the components of the graviphoton field,
see Table~\ref{grav}.  Observe that $T^{*+}=T^{-*}$. In this notation, the NGZ
identity (\ref{consistencyEqn}) takes the form
\begin{equation} 
\int d^{4}x\, \left[T^{*+}T^{+}- T^{*-}T^{-} \right]=0\, ,
\label{NGZsdcpx}
\end{equation}
In the linear Maxwell theory $T^{+}=0$, so  
\begin{equation} 
T^{+}= F^{+}- i G^{+}=0 \, ,
\end{equation}
which implies $\tilde{G}=-F$.  In more general theories in which the dual
field $G$ is treated as independent of $F$, this constraint is used to
eliminate the non-physical degrees of freedom and express $G$ as a function of
$F$ and scalars, if any, and it is known as a linear twisted self-duality
constraint.

In \cite{arXiv:1105.1273}, Bossard and Nicolai proposed to use a non-linear
deformation of the twisted self-duality constraint based on a manifestly
duality invariant source of deformation ${\cal I}^{(1)}(T)$ to construct a
self-dual theory. We will follow here the generalized procedure used in
\cite{Carrasco:2011jv}. Let us assume that a manifestly duality-invariant
${\cal I}^{(1)}(T)$ is given.  It was shown in \cite{Carrasco:2011jv} that if,
instead of vanishing as required by the linear twisted self-duality condition,
$T^{+}$ is given by the non-linear twisted self-duality condition
\begin{equation} 
\label{BNBI}
  T_{\mu\nu}^{+}  = \frac{\delta {\cal I}^{(1)}(T^{-},  T^{*+})}{\delta
    T^{*+}_{\mu\nu}} \, , 
  \qquad 
  (T_{\mu\nu}^{+})^{*}=  T_{\mu\nu}^{*-} = 
\frac{\delta {\cal I}^{(1)}(T^{-},  T^{*+})}{\delta  T^{-}_{\mu\nu}}\, , 
\end{equation}
it follows that the NGZ identity is satisfied automatically. One computes
$T^{*+}T^{+}- T^{*-}T^{-}$, using (\ref{BNBI}) and finds that it vanishes since it
is proportional to the variation of ${\cal I}^{(1)}(T)$ under duality, which
vanishes since $\delta\mathcal{I}^{(1)}=0$:
\begin{equation} 
\label{NGZsdcpx1}
  \int d^{4}x\, \left[T^{*+}T^{+}- T^{*-}T^{-}\right]=\int d^{4}x \left[T^{*+}
  \frac{\delta {\cal I}^{(1)}(T^{-},  T^{*+})}{\delta  T^{*+}} 
  - T^{-} \frac{\delta {\cal I}^{(1)}(T^{-},  T^{*+})}{\delta  T^{-}}\right]
={1\over B} \delta {\cal I}^{(1)}=0\, .
\end{equation} 
Thus, once the eqs.~(\ref{BNBI}) are solved for $G(F)$ there is no need to
check the NGZ identity, it is satisfied and we have the $G(F)$ of a self-dual
theory. 

In the models that we are going to study, the non-linear, twisted,
self-duality constraint can be solved as a power series in a parameter
$\lambda$:
\begin{equation}
\label{eq:Texpansion}
T^{+} = -2\sum_{n=1}^{\infty}\lambda^{n}T^{(n)+}\, ,  
\end{equation}
so
\begin{equation}
iG^{+}=F^{+} +2\sum_{n=1}^{\infty}\lambda^{n}T^{(n)+}\, , 
\end{equation}
from which we can get the coefficients of the series (\ref{eq:Gexpansion})
for $n>0$
\begin{equation}
\tilde{G}^{(n)} = -(T^{(n)+} +\mathrm{c.c.})\, ,\,\,\, (n>0)\, . 
\end{equation}
and of  (\ref{eq:Sexpansion}) for $n>0$ (for $n=0$ they have been chosen to
correspond to Maxwell's theory)
\begin{equation}
\label{eq:Sexpansion2}
2S^{(n)} = -\frac{1}{2(n+1)}\int d^{4}x\, 
\left[F^{+}T^{(n)+}+\mathrm{c.c.}\right]\, ,\,\,\, (n>0)\, . 
\end{equation}

\begin{table}[tp]%
\label{grav}\centering%
\begin{tabular*}{0.65\textwidth}{@{\extracolsep{\fill}} | c | c | c | }
  \hline
  Graviphoton  Components\ & Chirality & Charge  \\
  \hline
  $T^{+}=F^{+}-iG^{+}$\ & + & +  \\
  \hline
  $T^{*+}=F^{+}+i G^{+}$ & + & - \\
  \hline
  $T^{-}=F^{-}-iG^{-} $\ & -& + \\
  \hline
  $T^{*-}= F^{-}+i G^{-} $& -  &  -   \\
  \hline
\end{tabular*}
\caption{The 4 combinations of the graviphoton components have $\pm$ chirality and $\pm$ duality charge.}
\end{table}


\section{ Born Infeld  with higher derivatives and duality 
current conservation}

In this section we are going to construct two deformations of the Maxwell
theory using two particularly simple manifestly duality invariant sources of
deformation ${\cal I}^{(1)}_{A}(T)$ and ${\cal I}^{(1)}_{B}(T)$ given,
respectively, by 
\begin{eqnarray}
{\cal I}^{(1)}_{A}(T) 
& \equiv & 
\frac{\lambda}{2^{3}} 
t^{(8)}{}_{\mu_{1}\nu_{1}\mu_{2}\nu_{2}\mu_{3}\nu_{3}\mu_{4}\nu_{4}}
\partial_{\alpha}T^{*+\, \mu_{1}\nu_{1}} 
\partial^{\alpha}T^{-\, \mu_{2}\nu_{2}}
\partial_{\beta}T^{*+\, \mu_{3}\nu_{3}} 
\partial^{\beta}T^{-\, \mu_{4}\nu_{4}}\, ,
\\
& & \nonumber \\
{\cal I}^{(1)}_{B}(T) 
& \equiv & 
\frac{\lambda}{2^{3}} 
t^{(8)}{}_{\mu_{1}\nu_{1}\mu_{2}\nu_{2}\mu_{3}\nu_{3}\mu_{4}\nu_{4}}
\partial_{\alpha}T^{*+\, \mu_{1}\nu_{1}} 
\partial_{\beta}T^{-\, \mu_{2}\nu_{2}}
\partial^{\alpha}T^{*+\, \mu_{3}\nu_{3}} 
\partial^{\beta}T^{-\, \mu_{4}\nu_{4}}\, ,
\end{eqnarray}
where the tensor $t^{(8)}$ is defined in the Appendix, or, using the shorthand
notation introduced in the Appendix,
\begin{eqnarray}
{\cal I}^{(1)}_{A}(T) 
& \equiv & 
\frac{\lambda}{2^{3}} 
t^{(8)}{}_{abcd}
\partial_{\alpha}T^{*+\, a} 
\partial^{\alpha}T^{-\, b}
\partial_{\beta}T^{*+\, c}
\partial^{\beta}T^{-\, d}\, ,
\\
& & \nonumber \\
{\cal I}^{(1)}_{B}(T) 
& \equiv & 
\frac{\lambda}{2^{3}} 
t^{(8)}{}_{abcd}
\partial_{\alpha}T^{*+\, a} 
\partial_{\beta}T^{-\, b}
\partial^{\alpha}T^{*+\, c}
\partial^{\beta}T^{-\, d}\, .
\end{eqnarray}
At first order in $\lambda$ the models that one obtains using the procedure
described in the previous section are associated to the following deformations
of the action
\begin{eqnarray}
S_{A}^{(1)} 
& = &   
\tfrac{1}{4}\int d^{4}x\, 
t^{(8)}{}_{abcd}
\partial_{\alpha}F^{+\, a} 
\partial^{\alpha}F^{-\, b}
\partial_{\beta}F^{+\, c}
\partial^{\beta}F^{-\, d}\, ,
\\
& & \nonumber \\
S_{B}^{(1)} 
& = &   
\tfrac{1}{4}\int d^{4}x\, 
t^{(8)}{}_{abcd}
\partial_{\alpha}F^{+\, a} 
\partial_{\beta} F^{-\, b}
\partial^{\alpha}F^{+\, c}
\partial^{\beta}F^{-\, d}\, .
\end{eqnarray}
Alternative forms of these corrections which do not use the $t^{(8)}$ tensor
are eqs.~(\ref{eq:alt1}) and (\ref{eq:alt2}). 

In what follows we are going to construct explicitly the model A, using 
${\cal I}^{(1)}_{A}(T)$ in the non-linear twisted self-dual condition.


\subsection{Model A}

The simplest way to solve the non-linear twisted self-dual condition with
${\cal I}^{(1)}_{A}(T)$ is to plug the series expansion (\ref{eq:Texpansion})
into both sides of it and identify the terms with the same powers of
$\lambda$. First, observe that the expansion (\ref{eq:Texpansion}) for $T^{+}$
implies for $T^{*+}$ and $T^{-}$
\begin{eqnarray}
T^{*+} 
& = & 
2F^{+}+2\sum_{n=1}^{\infty}\lambda^{n}T^{(n)+}\, ,  
\\
& & \nonumber \\
T^{-} 
& = & 
2F^{-}+2\sum_{n=1}^{\infty}\lambda^{n}T^{(n)+*}\, .    
\end{eqnarray}
it is, then, convenient, to define\footnote{Notice, however, the expansion of
  $T^{+}$ is still given by (\ref{eq:Texpansion}) and has no term of zero
  order in $\lambda$.}
\begin{equation}
T^{(0)+} = F^{+}\, ,  
\end{equation}
so
\begin{eqnarray}
T^{*+} 
& = & 
2\sum_{n=0}^{\infty}\lambda^{n}T^{(n)+}\, ,  
\\
& & \nonumber \\
T^{-} 
& = & 
2\sum_{n=0}^{\infty}\lambda^{n}T^{(n)+*}\, .    
\end{eqnarray}
With these definitions, the non-linear twisted self-dual condition for this
model, which is
\begin{equation}
T^{+}_{a} = -\frac{\lambda}{2^{2}} t^{(8)}{}_{abcd}
\partial_{\alpha}(\partial^{\alpha}T^{-b}\partial_{\beta}T^{*+c}\partial^{\beta}T^{-d})\, ,
\end{equation}
takes the form
\begin{equation}
\sum_{n=1}^{\infty}\lambda^{n}T_{a}^{(n)+}
=
t^{(8)}{}_{abcd}
\sum_{p,q,r=0} \lambda^{p+q+r+1}
\partial_{\alpha}(\partial^{\alpha}T^{(p)-b}
\partial_{\beta}T^{(q)*+c}\partial^{\beta}T^{(r)-d})\, ,
\end{equation}
from which it follows that 
\begin{equation}
\label{eq:recursiverelation}
T_{a}^{(n)+}
=
t^{(8)}{}_{abcd}
\sum_{p,q,r=0}\delta_{p+q+r+1,n}
\partial_{\alpha}(\partial^{\alpha}T^{(p)-b}
\partial_{\beta}T^{(q)*+c}\partial^{\beta}T^{(r)-d})\, ,
\end{equation}
which can be solved recursively, given that $T^{(0)+}_{a}=F^{+}_{a}$. Thus, 
\begin{eqnarray}
T_{a}^{(1)+}
& = & 
t^{(8)}{}_{abcd}
\partial_{\alpha}(\partial^{\alpha}F^{-b}
\partial_{\beta}F^{+c}\partial^{\beta}F^{-d})\, ,
\\
& & \nonumber \\
T_{a}^{(2)+}
& = & 
t^{(8)}{}_{abcd}\left[
\partial_{\alpha}(\partial^{\alpha}F^{-b}
\partial_{\beta}F^{+c}\partial^{\beta}T^{(1)-d})
+
\partial_{\alpha}(\partial^{\alpha}F^{-b}
\partial_{\beta}T^{(1)+c}\partial^{\beta}F^{-d})
+
\partial_{\alpha}(\partial^{\alpha}T^{(1)-b}
\partial_{\beta}F^{+c}\partial^{\beta}F^{-d})
\right]\, ,
\\
& & \nonumber \\
T_{a}^{(3)+}
& = & 
t^{(8)}{}_{abcd}\left[
\partial_{\alpha}(\partial^{\alpha}F^{-b}
\partial_{\beta}F^{+c}\partial^{\beta}T^{(2)-d})
+
\partial_{\alpha}(\partial^{\alpha}F^{-b}
\partial_{\beta}T^{(1)+c}\partial^{\beta}T^{(1)-d})
+
\mathrm{permutations }
\right]\, ,
\end{eqnarray}
etc.  The action can be obtained immediately by using the power series
expansion of reconstructive identity (\ref{eq:Sexpansion2}). Explicitly, we
get:
\begin{eqnarray}
2S^{(0)}
&  = & 
-\tfrac{1}{4} \int \, d^{4}x \, F^{2}\, ,
\\
& & \nonumber \\
2S^{(1)} 
& = & 
\tfrac{1}{2}\int d^{4}x\, 
t^{(8)}{}_{abcd}
\partial_{\alpha}F^{+\, a} 
\partial^{\alpha}F^{-\, b}
\partial_{\beta}F^{+\, c}
\partial^{\beta}F^{-\, d}\, ,
\\
& & \nonumber \\
2S^{(2)} 
& = & 
-\tfrac{1}{2}\int d^{4}x\, 
\left[T^{(1)+}{}_{a}T^{(1)+\, a} +\mathrm{c.c.}\right]
\nonumber \\
& = &
-\tfrac{1}{2}\int d^{4}x\, 
\left\{t^{(8)}{}_{abcd}t^{(8)}{}_{defg}
\partial_{\alpha}\left(
\partial^{\alpha}F^{-\, b}
\partial_{\beta}F^{+\, c}
\partial^{\beta}F^{-\, d}\right)
\partial_{\gamma}\left(
\partial^{\gamma}F^{-\, e}
\partial_{\delta}F^{+\, f}
\partial^{\delta}F^{-\, g}\right)
+\mathrm{c.c.} \right\}
\, ,
\end{eqnarray}
etc., up to total derivatives.  It can be checked order by order that this
action is related to the dual field strength $iG^{+}=F^{+}-T^{+}$ by
(\ref{covConstraint}):
\begin{equation}
G^{+}{}_{\mu\nu} =2i \frac{\delta S}{\delta F^{+\, \mu\nu}}\, ,  
\end{equation}
as required.


\subsection{Model B}

The recursive algorithm for generating a complete action above produces the
$\lambda^{n}$ term from the previous ones. The derivation of this model
follows the exact steps which we outlined in the case A. Each time the
sequence of $(+ - + -)$ has to be replaced by $(+ + - -)$, the rest is the
same. Therefore we will not provide more details on the derivation of the B
model.


\section{Supersymmetrizable Born-Infeld duality 
symmetric model with higher derivatives}
 
The model with derivatives of $F$ known from the open superstring effective
action \cite{Print-88-0247 (LEBEDEV)} was shown to satisfy the NGZ current
conservation condition (\ref{consistencyEqn}) in \cite{Chemissany:2006qd}. In
this model the first deformation of the Maxwell theory is given by the quartic
coupling term
\begin{equation}
\label{S1susy}
S^{(1)}= 
\frac{\lambda}{2^{4}}\,  \int d^{4}x\, t^{(8)}_{abcd}
\partial_{\mu}F^{a}\partial^{\mu}F^{b}\partial^{\nu}F^{c}\partial_{\nu}F^{d}\, ,
\end{equation} 
in the notation introduced in the Appendix. As shown there, it can be
rewritten in the form (eq.~(\ref{eq:relation}))
\begin{equation}
\label{S1susy-2}
S^{(1)}= 
\frac{\lambda}{2^{2}}\int d^{4}x\, t^{(8)}{}_{abcd}\, 
\left[\, \partial_{\mu}F^{+a}\partial^{\mu}F^{-b}
\partial_{\nu}F^{+c}\partial^{\nu}F^{-d}
+\tfrac{1}{2} 
\partial_{\mu}F^{+a}\partial_{\nu}F^{-b}
\partial^{\mu}F^{+c}\partial^{\nu}F^{-d}
\, \right]\,  .
\end{equation} 

It is clear that, to reproduce this quartic term in the action we must take a
combination of the models A and B studied above and use, as manifestly
self-dual source of deformation 
\begin{equation}
\label{manifest4dF}
\mathcal{I}_{\rm string}^{(1)}(T)=
\mathcal{I}_{A}^{(1)}(T)+\tfrac{1}{2}\mathcal{I}_{B}^{(1)}(T)=
\frac{\lambda}{2^{3}}\,
t^{(8)}_{abcd}[\partial_{\mu}T^{*+\,a}\partial^{\mu}T^{-\,b}\partial_{\nu}
T^{*+\,c}\partial^{\nu}T^{-\,d}+\frac{1}{2}\partial_{\mu}T^{*+\,a}
\partial^{\mu}T^{*+\,b}\partial_{\nu}
T^{-\,c}\partial^{\nu}T^{-\,d}]\, .
\end{equation}
the resulting non-linear, twisted, self-duality constraint can be solved by
the same recursive procedure we employed for the model A above and the dual
field strength $G(F)$ and corresponding action can be found by the use of the
new reconstructive identity.

It is interesting to observe that, after partial integration, the above
quartic term is very close to the $(\partial F)^{4}$ term found in
ref.~\cite{arXiv:1101.1672}, although the latter, corresponding to an
amplitude calculation, is not real in its current form. It is likely that for
the effective action one can produce the real expression dividing the one in
ref.~\cite{arXiv:1101.1672} by two and adding the Hermitean conjugate.


\section{More general $U(1)$ duality, no scalars models}

Using the covariant procedures for perturbative non-linear deformation of
duality-invariant theories \cite{Carrasco:2011jv} we can construct more
general models with NGZ current conservation. For example, we may consider
more general sources of deformation.
\begin{equation} 
\mathcal{I}^{(1)}_{f_n}(T^{-},T^{*+})= 
\sum_{n=1}  f_n \Big (\mathcal{I}^{(1)}(T^{-},T^{*+})\Big )^{n} \ ,
\end{equation}
where $\mathcal{I}^{(1)}(T^{-},T^{*+})$ is defined in (\ref{manifest4dF}) and
$f_n$ are arbitrary constants, and the model we described above in details has
$f_1=1$ and $f_n=0, n>1$.  In addition, we may add terms which depend only on
$F$'s without derivatives, for example, the ones studied in
\cite{Carrasco:2011jv}.

Any manifestly $U(1)$ duality invariant $\mathcal{I}(T^{-},T^{*+})$ with
space-time derivatives action of $T$'s or without, has to have the same number
of $T^{-}$' s as $T^{*+}$'s, and has to be Lorentz covariant. In such case,
one expects a recursive equation, defining $G_{\mu\nu}(F)$ from the equation
\begin{equation} 
\label{CKR}
T_{\mu\nu}^{+}  = \frac{\delta {\cal I}(T^{-},  T^{*+})}{\delta  T^{*+}_{\mu\nu}}\, ,
\end{equation}
as shown in the simple models defined in \cite{Carrasco:2011jv} without
derivatives and in sec. III in case with derivatives.  The solution is
guaranteed to satisfy the NGZ $U(1)$ current conservation
\cite{Carrasco:2011jv}.

This eq.~(\ref{CKR}) in all cases provides a recursive procedure determining
$G(F)$ as a powers series in $\lambda$. The action in these most general
models of $U(1)$ duality without scalars is given by the new reconstructive
identity (\ref{new}).


\section{Discussion}

In this paper we have constructed explicitly the first complete model of
Born-Infeld type with higher derivatives, which has an electromagnetic $U(1)$
duality. The model is given by the power series expansion of the Lagrangian
and involves all powers of the Maxwell field strength and their derivatives.
\begin{equation}
\label{Action1}
S= S_{\rm Maxwell} 
-\sum_{n=1}^{\infty} \frac{\lambda^{n}}{2 (n+1)} \int d^{4}x\, 
\big[F^{+\mu\nu}T^{(n)+}{}_{\mu\nu}+\mathrm{c.c.}\big] \ .
\end{equation}
Here $T^{(n)}\sim \partial^{4n} F^{2n+1}$ and the explicit expression is
given via a recursive algorithm in eq.~(\ref{eq:recursiverelation}), which
defines $T^{(n)}$ in terms of $T^{(m)}$ with $m<n$ and starts with $T^{(0)+}=
F^{+}$.  We have also outlined the procedure to produce more complicated
models where terms with and without derivatives on $F$ are mixed.

Apart from the intrinsic motivation to discover a non-linear model with higher
derivatives and with duality symmetry, which was not known in the past, our
goal here was to test the Bossard-Nicolai proposal \cite{arXiv:1105.1273}. The
authors conjectured that there is a straightforward algorithm which allows to
construct $N=8$ supergravity with higher derivatives, consistent with
$E_{7(7)}$ duality. This conjecture was used in \cite{arXiv:1105.1273} to
counter the argument of \cite{Kallosh:2011dp} suggesting that $E_{7(7)}$
duality symmetry predicts the finiteness of ${\cal N}=8$ supergravity.

However, there is no actual construction of ${\cal N}=8$ supergravity with
higher derivatives in \cite{arXiv:1105.1273}, which would be a formidable
task. Therefore we performed a detailed investigation of this issue in
applications to much simpler models, such as the Born-Infeld models and their
generalizations.  An investigation of this issue in \cite{Carrasco:2011jv}
demonstrated that the algorithm of construction of ${\cal N}=8$ supergravity
with higher derivatives requires substantial modifications even in application
to the simplest Born-Infeld model.  Moreover, it was observed in
\cite{Carrasco:2011jv} that the presence of the 4-point UV divergence $ F^{4}
f(s, t, u) $ term in the ${\cal N}=8$ supergravity would require to produce a
theory of the Born-Infeld type with derivatives leads to a non-stop
proliferation of the powers of the vector field strength with increasing
number of derivatives.

A generalization of the results in \cite{Carrasco:2011jv} to the Born-Infeld
model with higher derivatives required additional efforts. In this paper we
were able to construct a toy model of a Born-Infeld ${\cal N}=8$ supergravity,
with ${\cal N}=0$ supersymmetry replacing ${\cal N}=8$ and $U(1)$ duality
replacing $E_{7(7)}$. The model indeed has a full non-linearity in powers of
$\lambda^{n} \partial ^{4n} F^{2n+2} $ with $n\rightarrow \infty$, as
predicted in \cite{Carrasco:2011jv}.

Thus, whereas we were able to construct the Born-Infeld model with
derivatives, and we are planning to develop a similar construction for
supersymmetric models, which also have a $U(1)$ duality symmetry, at present
we do not see any obvious way to extend this construction and develop the
Born-Infeld version of ${\cal N}=8$ supergravity along the lines of
\cite{arXiv:1105.1273}.  Until the existence of the Born-Infeld version of
${\cal N}=8$ supergravity is demonstrated, the argument that $E_{7(7)}$
duality symmetry predicts the finiteness of ${\cal N}=8$ supergravity
\cite{Kallosh:2011dp} seems  still valid. Moreover, even if one manages to
construct a consistent 2-coupling model of ${\cal N}=8$ supergravity with
gravitational coupling $\kappa^2$ as well as Born-Infeld coupling $\lambda$,
it will raise a new question whether the conjectured existence of this new
theory predicts anything for the UV behaviour of the original one-coupling
${\cal N}=8$ supergravity, which depends only on gravitational coupling.


\section*{Acknowledgments}

We are grateful to G.~Bossard, J.~Broedel, J.J.~Carrasco, S.~Ferrara,
D.~Freedman, M. Green, A.~Linde, H.~Nicolai, R.~Roiban, E. Silverstein and
A. Tseytlin for stimulating discussions and especially to M.~de Roo for his
contribution in the early stages of this project.  This work is supported by
the Stanford Institute for Theoretical Physics and the NSF grants 0756174, the
Spanish Ministry of Science and Education grant FPA2009-07692, the Comunidad
de Madrid grant HEPHACOS S2009ESP-1473, and the Spanish Consolider-Ingenio
2010 program CPAN CSD2007-00042.  W.C.~and T.O.~wish to thank the Stanford
Institute for Theoretical Physics for its hospitality and financial support.

\appendix

\section{Some useful relations}
\label{appendix}

Using the definition of the Hodge dual 
\begin{equation}
\tilde{A}_{\mu\nu}\equiv
\tfrac{1}{2}\epsilon_{\mu\nu\rho\sigma}A^{\rho\sigma}\, ,
\hspace{1cm}
\tilde{\tilde{A}}=-A\, ,  
\end{equation}
and of the self- and anti-self-dual parts
\begin{equation}
A^{\pm} = \tfrac{1}{2}(A\pm i\tilde{A})\, ,  
\end{equation}
for 2-forms, one can easily prove the following identities involving arbitrary
2-forms $A$ and $B$:
\begin{eqnarray}
\tilde{A}\tilde{B}
& = &
BA-\tfrac{1}{2}\textrm{tr}(AB)\mathbbm{1}\, ,\\
\nonumber \\
\tilde{A} B 
& = & 
-\tilde{B}A+\tfrac{1}{2}\textrm{Tr}(A\tilde{B})\mathbbm{1}\, ,\\
 \nonumber \\
A^{\pm}B^{\pm} 
& = & 
-B^{\pm}A^{\pm}+\tfrac{1}{2}\textrm{Tr} (A^{\pm}B^{\pm})\mathbbm{1}\, ,\\
\nonumber \\
A^{\pm}B^{\mp}
& = & B^{\mp}A^{\pm}\, .
\end{eqnarray}

The $t^{(8)}$ tensor \cite{CALT-68-911} is totally symmetric in four pairs of
antisymmetric indices. It is convenient to use only one Latin index
$a,b,c\ldots$ to denote each of these four pairs and write $t^{(8)}{}_{abcd}$
instead of
$t^{(8)}{}_{\mu_{1}\nu_{1}\mu_{2}\nu_{2}\mu_{3}\nu_{3}\mu_{4}\nu_{4}}=
t^{(8)}{}_{[\mu_{1}\nu_{1}][\mu_{2}\nu_{2}][\mu_{3}\nu_{3}][\mu_{4}\nu_{4}]}$.
Then, in terms of these indices, $t^{(8)}$ is completely symmetric
$t^{(8)}{}_{abcd}= t^{(8)}{}_{(abcd)}$.

$t^{(8)}$ can be defined by its contraction with 4 arbitrary 2-forms $A,B,C,D$:
\begin{eqnarray}
t^{(8)}{}_{abcd}A^{a}B^{b}C^{c}D^{d}
& = & 
8[\textrm{Tr}(ABCD)
+\textrm{Tr}(ACBD)
+\textrm{Tr}(ACDB)]
\nonumber\\
\nonumber \\
& &
-2[ \textrm{Tr}(AB)\textrm{Tr}(CD)
+\textrm{Tr}(AC)\textrm{Tr}(BD)
+\textrm{Tr}(AD)\textrm{Tr}(BC)]\, .
\end{eqnarray}

Then, using the above relations, one can write
\begin{equation}
t^{(8)}{}_{ abcd}\, 
\partial_{\mu}F^{a}\partial^{\mu}F^{b}\partial_{\nu}F^{c}\partial^{\nu} F^{d}
 =  
16\, \left\{ 
\textrm{Tr}\left(\partial_{\mu}F^{+}\partial_{\nu}F^{+}\right)
\textrm{Tr}\left(\partial^{\mu}F^{-}\partial^{\nu}F^{-}\right)
+\tfrac{1}{2}\textrm{Tr}\left(\partial_{\mu}F^{+}\partial^{\mu}F^{+}\right)
\textrm{Tr}\left(\partial_{\nu}F^{-}\partial^{\nu}F^{-}\right)
\right\}\, ,
\end{equation}
and
\begin{eqnarray}
\label{eq:alt1}
t^{(8)}{}_{abcd}\, 
\partial_{\mu}F^{+a}\partial^{\mu}F^{-b}
\partial_{\nu}F^{+c}\partial^{\nu}F^{-d}
& = & 
4\textrm{Tr}\left(\partial_{\mu}F^{+}\partial_{\nu}F^{+}\right)
\textrm{Tr}\left(\partial^{\mu}F^{-}\partial^{\nu}F^{-}\right)\, ,
\\
\nonumber \\
\label{eq:alt2}
t^{(8)}{}_{abcd}\, 
\partial_{\mu}F^{+a}\partial_{\nu}F^{-b}
\partial^{\mu}F^{+c}\partial^{\nu}F^{-d}
& = & 
4 \textrm{Tr}\left(\partial_{\mu}F^{+}\partial^{\mu}F^{+}\right) 
\textrm{Tr}\left(\partial_{\nu}F^{-}\partial^{\nu}F^{-}\right)\, ,
\end{eqnarray}
from which we find that 
\begin{equation}
\label{eq:relation}
t^{(8)}{}_{abcd}\, 
\partial_{\mu}F^{a}\partial^{\mu}F^{b}\partial_{\nu}F^{c}\partial^{d}
= 
4t^{(8)}{}_{abcd}\, 
\left[\, \partial_{\mu}F^{+a}\partial^{\mu}F^{-b}
\partial_{\nu}F^{+c}\partial^{\nu}F^{-d}
+\tfrac{1}{2} 
\partial_{\mu}F^{+a}\partial_{\nu}F^{-b}
\partial^{\mu}F^{+c}\partial^{\nu}F^{-d}
\, \right]\,  .
\end{equation}


\end{document}